# DISCUSSION OF: STATISTICAL ANALYSIS OF AN ARCHEOLOGICAL FIND

By J. Mortera and P. Vicard

*Università Roma Tre*

**1. Introduction.** The paper by Feuerverger analyses interesting data on the inscriptions found on the ossuaries of a burial tomb unearthed in Jerusalem in 1980. A statistical analysis is made of the plausibility that the names inscribed on the ossuaries match those of the New Testament (NT) figures. The evidence on which the analysis is based is the distribution of names in the era when the tomb was dated. The results are based on assumptions which may drive some of the results.

Some questions immediately come to mind.

- The author assumes that a tomb of Jesus of Nazareth exists—this assumption is disputed by many people, as stated by Colin Aitken in the interview given on March 1, 2007 to *The Herald*. Moreover, even assuming the existence of a tomb of Jesus of Nazareth, why should it be located in Talpiyot and not, say, at the Sepulchre in Jerusalem or in another site or city?
- What is the uncertainty of the estimated number 1,100 of inscribed adult ossuaries? It would be important to measure the variability around that estimate.
- What implications does the statement that the Talpiyot finding is the "best of many trials" have on the results?
- Why was the DNA evidence available only for the ossuaries with the inscriptions "Yeshua son of Yhosef" and "Mariamenou e Mara?" Why was DNA not extracted from all the remains?
- Assumption A.7, which interprets the name on Ossuary #1 as being that of Mary Magdelene, is one factor that has a very strong influence on the results of the analysis since it is such a rare name. Is there no uncertainty in this interpretation?

---

Here we discuss further aspects of the paper and propose possible ways in which the statistical analysis could be extended.

The assumptions made by the author are based both on *anonymous* sources, such as the 4th century CE version of the Acts of Philip[1] and the NT gospels written between 65 and 100 CE. A possible way to handle the different reliability of these sources could have been that of assigning different weights to the assumptions based on historical sources and to those based on other sources, such as the apocryphal narratives.

Since a hypothesis such as the one investigated by the author could have an impact on the history of religion, it would be appropriate to examine other pieces of evidence. These could help explore the plausibility that the Talpiyot family configuration was so rare at that time that there could have been only one family with that configuration.

We will base our discussion on the following issues: in Section 2 we show how to deal with the uncertainty in name frequencies; comments on the DNA evidence are given in Section 3; the analysis of different items of evidence is given in Section 4 and Section 5 shows how an object-oriented Bayesian network (OOBN) can be structured for combining different items of evidence.

**2. Uncertain name frequencies.** In Section 5 the author gives details on the available documentation that could be used to obtain the distribution of names in the era relevant to the study. The name frequencies of three different sources are shown. Table 1 (from Table 2 in the paper) shows the relative frequency of Ilan's nonossuary and ossuary names. Category "Other" indicates all the other names having overall frequency $f_i = 1 - \sum_j f_j$.

The author tells us that "the relative frequency of female ossuaries (names) is under represented" since sometimes fathers (and occasionally husbands) were named on female ossuaries. Furthermore, the name distribution sources refer to a range in time period wider than that of the burial tomb in question. There is thus potential bias and many sources of uncertainty in the name frequency distributions. This should be appropriately accounted for, not by ad hoc adjustments, but in a fully probabilistic framework.

Thus, when analyzing the data, the name frequencies are not fixed probabilities, but empirical frequencies. These are most probably *not* a random sample from the population of names of the era. The uncertainty about these name frequencies can be modeled by assuming a Dirichlet prior and multinomial sampling. In Green and Mortera (2008) we show how to model uncertain frequency distributions in forensic inference in a fully probabilistic way in a Bayesian network [Cowell et al. (1999)]. Taking all uncertainties into account, in a probabilistically coherent way, would avoid those arbitrary adjustments (like multiplying by 5 or dividing by 1.2) that are made in computing the $RR$ values.

---

[1] Craig A. Blaising, "Philip, Apostle." In *The Encyclopedia of Early Christianity*, Everett Ferguson, ed. (New York: Garland Publishing, 1997).

Furthermore, a very strong assumption made is that of considering independence among the names and then applying the product rule to obtain the overall $RR$ value. Also, the fact that brothers do not commonly have the same name is ignored. These dependencies as well as the fact that "in assignment of names within a family, children frequently are named as earlier 'nodes' in the family tree" can be taken into account in structuring a Bayesian network to analyze this problem.

Finally, *all* uncertainties, the name frequency distributions, the number of inscribed adult ossuaries and the relevant population size should be accounted for and modeled appropriately.

**3. DNA evidence.** The discriminatory power of DNA analysis in forensic identification is well known. Mitochondrial (mtDNA), Y-chromosome DNA and even nuclear DNA can be extracted from ancient human remains. This information is extremely important for reconstructing a probable family pedigree and establishing the sex of the owners of the bones. From this analysis one can compute the probability that the bones either belong to individuals of the same nuclear family, or to possible relatives of the family, or are from unrelated individuals. So, as stated before, why was the mtDNA of the bones found only in the ossuaries with the inscriptions "Yeshua son of Yhosef" and "Mariamenou e Mara" analyzed?

TABLE 1
*Frequency distribution of Jewish female names*

| Names | Ilan nonossuary | Ilan ossuaries |
|---|---|---|
| Mary | 0.242 | 0.228 |
| Salome | 0.161 | 0.212 |
| Shelamzon | 0.048 | 0.098 |
| Martha | 0.032 | 0.088 |
| Joanna | 0.040 | 0.036 |
| Shiphra | 0.024 | 0.047 |
| Berenice | 0.056 | 0.010 |
| Sara | 0.024 | 0.026 |
| Imma | 0.016 | 0.031 |
| Mara | 0.016 | 0.026 |
| Other | 0.339 | 0.197 |
| N. females | 317 | 193 |

In the well-known Romanov case, mtDNA played a central role in the attempt to discover whether Anastasia, the daughter of the Tsar Nicholas II, was killed and buried with her parents [Gill et al. (1994)]. Nine skeletons unearthed in Ekaterinburg, Russia, in 1991, were tentatively identified as the remains of the last Tsar, his family and the Royal Physician and three servants. Sex testing and nuclear DNA were extracted from the bones in order to confirm that a family group was present in the grave. mtDNA (and Y-chromosome DNA) is transmitted unchanged—apart from the possibility of mutations—in the maternal (paternal) line. To verify the hypothesis that these remains were effectively from the Tsar, the Tsarina and their children, the DNA of their living descendants were analyzed, among which that of the Duke of Edinburgh. The DNA evidence supported the hypothesis that the remains were those of the Romanov family. From all the evidence—the DNA analysis, the statistical analysis and historical facts—the conclusion was reached that the nine skeletons were those of Tzar Nicolas II, the Tzarina, three of their four daughters, the court doctor and three servants. A complex statistical analysis was also made to obtain the most probable pedigree given the DNA evidence [Egeland et al. (2000)].

Although the Romanov remains are of much more recent origin than the bones found in the Jerusalem ossuaries, DNA can be extracted from ancient remains. In fact, both mtDNA and nuclear DNA has been extracted from fossils of a Neandertal man [Green et al. (2006)].

In contrast to the Romanov case, we do not have *known* descendants of the NT family. Therefore, the DNA analysis can only be used to verify the hypothesis about a specific pedigree. It can thus help to disconfirm the hypotheses that this is the NT family, but cannot be used to confirm that the hypothesis is true.

Furthermore, information on the dating and measurements taken from the ossuaries and the human remains, would be helpful to determine the age group, sex and estimated burial time of each remain.

**4. Analyzing many items of evidence.** There are many similarities in the analysis made in this paper to those commonly made in forensic identification, some of which we will illustrate here. Figure 1 shows a pictorial representation of a network for analyzing two different items of evidence pertaining to the hypotheses of interest. In this case, it is not possible to make forensic identification but it is only possible to make inference about specific pedigrees.

Let $\mathcal{E}$ denote one or more items of evidence (perhaps the totality). We need to consider how this evidence affects the comparison of the hypotheses, $H_0$:Tomb=NTped, the tombsite belonged to a family with a pedigree like that of the NT family;[2] one alternative hypothesis could be $H_1$ : Tomb$\neq$NTped, the tombsite does *not* belong to a family with a pedigree equal to that of the NT family. This alternative hypothesis could be formed by a number of hypotheses pertaining to each possible relationship.

When we are only comparing two hypotheses $H_0$ and $H_1$, the impact of the totality of say $k$ different elements of evidence $\mathcal{E} = (\mathcal{E}_1, \ldots, \mathcal{E}_k)$, from all sources, is embodied in the *likelihood ratio*,

$$(1) \qquad LR = P(\mathcal{E}|H_1)/P(\mathcal{E}|H_0).$$

When the items of evidence $\mathcal{E}_i$ for $i = 1, \ldots, k$ are conditionally independent given the hypotheses, the overall $LR$ can be computed as $LR = \prod_i LR_i$, where $LR_i = P(\mathcal{E}_i|H_1)/P(\mathcal{E}_i|H_0)$. Given the likelihood ratio, $LR_i$, based on the distribution of names (loosely, onomasticon) this can be updated with the $LR$s based on other items of evidence (e.g., all DNA profiles) and the evidence given in (1) to (10) of Section 14, to form the overall likelihood ratio.

We thus do not see the reason why the author excludes the possibility of computing a $LR$ and of using other pieces of evidence as well.

**5. OOBN for analyzing two or more pieces of evidence.** An object-oriented Bayesian network for analysing two or more pieces of evidence. OOBNs have shown to be an extremely versatile tool to handle different pieces of evidence relating to an identification issue; see, among others, Cowell, Lauritzen and Mortera (2007), Dawid, Mortera and Vicard (2007) and Taroni et al. (2006). A network can be built to compute the overall likelihood ratio given all the pieces of evidence.

Figure 1 shows an example of an OOBN for evaluating the weight of two pieces of identification inference: that from onomasticon together with that from DNA profiling.

In the network, the two hypotheses, described in Section 4, bearing on the pedigree of the tombsite ownership, are represented by the *true/false* states of the Boolean node Tomb=NTped?. The onomasticon node represents a complex subnetwork having as input both the Female and Male name frequencies, represented by nodes F name frequency and M name frequency, respectively. For example, the probability distribution and states of node F name frequency are given in Table 1. The DNA node represents another complex subnetwork having as input the gene frequencies represented by nodes gene frequency. The evidence on the tombstone names and the DNA extracted from the bones is entered in onomasticon and DNA and propagated throughout the entire network yielding, in node Tomb=NTped?, the overall likelihood ratio based on *all* the evidence.

---

[2]The fact that no official sources contain information about Jesus from Nazareth having had sons should be appropriately considered.

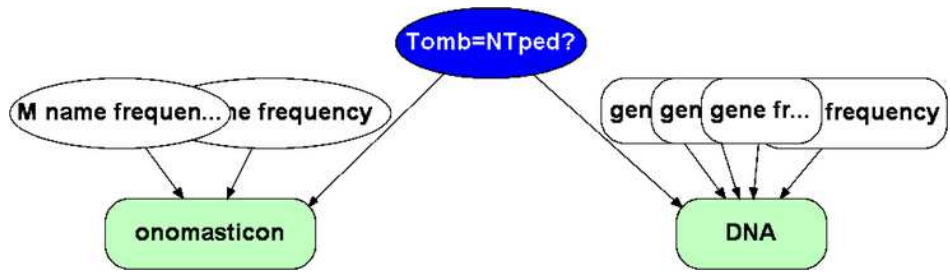

Fig. 1. *OOBN for tomb identification using onomasticon and DNA evidence.*

We enjoyed reading the paper and writing this discussion. We recognize that Feuerverger does not have the DNA test results, but we wonder if he could facilitate access to these data so that further analysis could be made on this interesting case.

Dipartimento di Economia
Università Roma Tre
Via Silvio D'Amico 77
00145 Roma
Italy
E-mail: mortera@uniroma3.it
        vicard@uniroma3.it